\documentclass[12pt]{article}

\newcommand{\beq}{\begin{equation}}
\newcommand{\eeq}{\end{equation}}
\newcommand{\beqa}{\begin{eqnarray}}
\newcommand{\eeqa}{\end{eqnarray}}
\newcommand{\beqar}{\begin{eqnarray*}}
\newcommand{\eeqar}{\end{eqnarray*}}

\newcommand{\inn}{\!\cdot\!}

\newcommand{\la}{\lambda}
\newcommand{\Lam}{\Lambda}

\newcommand{\ie}{{\it i.e.,}\ }
\newcommand{\labell}[1]{\label{#1}} 
\newcommand{\reef}[1]{(\ref{#1})}
\newcommand\prt{\partial}

\newcommand\hD{\hat{D}}

\newcommand\tD{{\tilde D}}

\newcommand\tQ{{\tilde Q}}
\newcommand\tprt{{\tilde \partial}}

\newcommand\tT{{\tilde T}}

\newcommand\Tr{{\rm Tr}}
\newcommand\STr{{\rm STr}}

\begin{document}

 \vspace*{1cm}

\begin{center}
{\bf \Large
A proposal for  
M2-brane-anti-M2-brane  action

 }
\vspace*{1cm}

{Mohammad R. Garousi}\\
\vspace*{0.2cm}
{ Department of Physics, Ferdowsi University of Mashhad,\\ P.O. Box 1436, Mashhad, Iran}\\
\vspace*{0.1cm}
\vspace*{0.4cm}

\vspace{2cm}
ABSTRACT
\end{center}
We propose a manifestly $SO(8)$ invariant BF type Lagrangian for describing the dynamics of  M2-brane-anti-M2-brane system in flat spacetime. When one of the scalars which satisfies a free-scalar equation takes a large expectation value, the M2-brane-anti-M2-brane action reduces to the tachyon DBI  action of D2-brane-anti-D2-brane system in flat spacetime.  
\vfill \setcounter{page}{0} \setcounter{footnote}{0}
\newpage

\section{Introduction} \label{intro}
Following the idea that the Chern-Simons gauge theory may be used to describe the dynamics of coincident M2-branes \cite{Schwarz:2004yj}, Bagger and Lambert \cite{Bagger:2007vi} as well as Gustavsson \cite{Gustavsson:2007vu} have constructed three dimensional ${\mathcal{N}}=8$ superconformal $SO(4)$ Chern-Simons gauge theory based on 3-algebra.
It is believed that the BLG world volume theory at level one describes two M2-branes on $R^8/Z_2$ orbifold \cite{Lambert:2008et}. The world volume theory of N M2-branes on $R^8/Z_k$ orbifold has been constructed in \cite{Aharony:2008ug} which is given by ${\mathcal{N}}=6$ superconformal  $U(N)_k\times U(N)_{-k}$ Chern-Simons gauge theory.

The signature of the metric on 3-algebra in the BLG model is positive definite. This assumption has been relaxed in \cite{Gomis:2008uv} to study $N$ coincident M2-branes  in flat spacetime. The so called BF membrane theory with arbitrary semi-simple Lie group has been proposed in \cite{Gomis:2008uv}. This theory has ghost fields, however, there are different arguments that model may be unitary due to the particular form of the interactions \cite{Gomis:2008uv,Cecotti:2008qs}.  The bosonic part of the Lagrangian for gauge group $U(N)$ is given by
\beqa
L=\Tr\left(\frac{1}{2}\epsilon^{abc}B_aF_{bc}-\frac{1}{2}\hD_aX^I\hD^aX^I+\frac{1}{12}M^{IJK}M^{IJK}\right)+(\prt_aX^I_--\Tr(B_aX^I))\prt^aX^I_+\labell{L1}
\eeqa
where $A_a,B_a,X^I$ are in adjoint representation of $U(N)$ and $X^I_-,X^I_+$ are singlet under $U(N)$, and
\beqa
M^{IJK}&\equiv&X^I_+[X^J,X^K]+X^J_+[X^K,X^I]+X^K_+[X^I,X^J]\nonumber\\
\hD_aX^I&=&D_aX^I-X^I_+B_a\,,\qquad D_aX^I=\prt_aX^I-i[A_a,X^I]
\eeqa
Obviously the above Lagrangian  is invariant under global $SO(8)$ transformation  and under $U(N)$ gauge transformation associated with the $A_a$ gauge field. It is also invariant under gauge transformation associated with the $B_a$ gauge field
\beqa
\delta_B X^I=X^I_+\Lam\,,\qquad \delta_B B_a=D_a\Lam\,,\qquad \delta _BX^I_+=0\,,\qquad \delta _BX^I_-=\Tr(X^I\Lam)\labell{Bt}
\eeqa
The Lagrangian \reef{L1} is a candidate to describe the dynamics of N stable M2-branes in flat supergravity background. A nonlinear extension of this Lagrangian in nonabelian case is proposed in \cite{Iengo:2008cq,Garousi:2008xn} (see also \cite{Li:2008ya,Alishahiha:2008rs})\footnote{ Nonlinear action of M2-brane in the presence of background fields for abelian case has been discussed in \cite{Kluson:2008nw}.}.

In this paper, we would like to study  the dynamics of unstable M2-brane-anti-M2-brane system. The instability of this system can be either unperturbative effect or it can be  the result of  having tachyon fields in the spectrum of M2-brane-anti-M2-brane system, as in the D2-brane-anti-D2-brane system. Assuming the latter  case, one  may then use the Higgs mechanism  \cite{Mukhi:2008ux} to  find the effective action by including appropriately   the tachyons in the nonlinear action \cite{Iengo:2008cq,Garousi:2008xn}. That is, when one of the scalars $X^I_+$ takes a large expectation value,  M2-brane-anti-M2-brane action should be reduced to  the D2-brane-anti-D2-brane action. However, this mechanism does not work for the tachyon potential because the M2-brane-anti-M2-brane action should describe the D2-brane-anti-D2-brane system at strong coupling. One expects the tachyon potential at the strong coupling to be totally different than the tachyon potential at the weak coupling. So the Higgs mechanism can not fix the tachyon potential in terms of the tachyon potential of D2-brane-anti-D2-brane system. To find the M2-brane-anti-M2-brane action we do as follows: Near the unstable point, one can set the tachyon potential to one, and find the other  parts of  the M2-brane-anti-M2-brane action  by the Higgs mechanism. Then one multiplies  the result by the unknown M2-brane-anti-M2-brane tachyon potential.


In the next section we review the construction of the effective action of $D_2\bar{D}_2$ system  proposed in \cite{Garousi:2007fn}  which is a  nonabelian extension of the tachyon DBI action. Then we use de Wit-Herger-Samtleben duality transformation to write the $D_2\bar{D}_2$ action in a BF theory. In section 3, we propose an $SO(8)$ invariant BF type action for $M_2\bar{M}_2$ system which reduces to the above  theory  when one of the scalars $X^I_+$ takes a large expectation value. 

\section{D2-brane-anti-D2-brane effective action}

An effective action for $D_9\bar{D}_9$ system has been proposed in \cite{Sen:2003tm} whose vortex solution satisfies some consistency  conditions. This action has been written  as a non-abelian extension of the tachyon DBI action in \cite{Garousi:2007fn}. However, the ordering of the matrices in the action is not consistent with the S-matrix elements. Hence, another effective action has been proposed in \cite{Garousi:2007fn} which is consistent with the S-matrix elements. This second action may be related to the action proposed in \cite{Sen:2003tm} by some field redefinition. In the following we are going to review this second construction of the effective action for $D_2\bar{D}_2$ system.  

The   effective action for describing the 
dynamics of one non-BPS D$_p$-brane in flat background in static gauge is given by \cite{Sen:1999md,Garousi:2000tr,Bergshoeff:2000dq,Kluson:2000iy}:
 \beqa
S&=&-T_p\int d^{p+1}\sigma V(T^2)
\sqrt{-\det(\eta_{ab}+\prt_aX^i\prt_bX^i+\la F_{ab}+\la\prt_a T\prt_b T)} \,\,,\labell{dbiac2}\eeqa where $\la\equiv 2\pi\alpha'$ and
$V=1-\frac{\pi}{2}T^2+O(T^4)$ is the tachyon potential\footnote{Our index convention is that
$\mu,\nu,...=0,1,...,9$; $a,b,...=0,1,...,p$;  
$i,j,...=p+1,...,9$ and $I,J,...=3,4,..., 10$.}.  The action for $N$ non-BPS $D_p$-branes  may be given  by  some non-abelian extension of the above action. To study the non-abelian extension of the above action for arbitrary  $p$, one may first consider  the non-abelian action for $p=9$ case which has no transverse scalar field,  and then use the T-duality transformations to find the non-abelian  action for any $p$. 

The following  non-abelian action has been proposed in \cite{Garousi:2000tr} for describing the dynamics of $N$ non-PBS $D_9$-brames:
\beqa
S&\!\!\!=\!\!\!&-T_9\STr\int d^{10}\sigma V(T^2)\sqrt{-\det(\eta_{\mu\nu}+\la F_{\mu\nu}+
\la D_{\mu} T D_{\nu}T}) \labell{dbiac22}\eeqa 
where the symmetric trace make the integrand to be a Hermitian matrix. In above,  the gauge field
strength and covariant derivative of the
tachyon are\beqa
F_{\mu\nu}&=&\prt_{\mu}A_{\nu}-\prt_{\nu}A_{\mu}-i[A_{\mu},A_{\nu}]\,,\nonumber\\
D_{\mu}T&=&\prt_{\mu}T-i[A_{\mu},T]\,.\nonumber\eeqa Obviously the  action
 \reef{dbiac22} has $U(N)$ gauge symmetry and reduce to \reef{dbiac2} for $N=1$. 

The trace in the non-abelian action  \reef{dbiac22} is the symmetric trace. That is,  if one expands  the square root and the tachyon potential, then  the non-abelian expressions
of the form $F_{\mu\nu}, \,D_{\mu}T$ and  the individual
$T$ of the tachyon potential must appear in each term of the expansion as symmetric. This property make it possible to treat the non-abelian expressions
  $F_{\mu\nu}, \,D_{\mu}T$ and  
$T$  as ordinary number when manipulating them. 
  Various couplings in the action \reef{dbiac22} are consistent with the  appropriate disk level S-matrix elements in string
theory \cite{Garousi:2000tr,Garousi:2002wq,BitaghsirFadafan:2006cj}. In particular, the calculation in \cite{BitaghsirFadafan:2006cj} shows that the consistency is hold only if one uses the symmetric trace prescription. 

Using the effective action of $N$ non-BPS D$_9$-branes \reef{dbiac22}, one finds the effective action of $N$ non-BPS D$_2$-branes by using T-duality \cite{Garousi:2000tr}. 
The proposal for the effective action of  $D_2\bar{D}_2$ \cite{Garousi:2004rd,Garousi:2007fn} is then to project the effective action of $N=2$ non-BPS D$_2$-branes  with $(-1)^{F_L}$, \ie the matrices $A_a$, $X^i$ and $T$ take the following form:
\beqa A_{a}=\pmatrix{A_{a}^{(1)}&0\cr
0&A_{a}^{(2)}},\,\,X^i=\pmatrix{X^{i(1)}&0\cr
0&X^{i(2)}},\,\,T=\pmatrix{0&\tau\cr \tau^*&0} \labell{M11}
\eeqa 
which reduces the $U(2)$ gauge symmetry to $U(1)\times U(1)$ gauge symmetry. 

Replacing  the above matrices in the effective action of $N=2$ non-BPS D$_2$-branes \cite{Garousi:2000tr}, one finds that the effective action of $D_2\bar{D}_2$  takes the following form:
 \beqa
S^{D\bar{D}}&=&-T_2\int
d^{3}\sigma \STr\left(V\sqrt{\det(Q)}\right.\labell{nonab}\\
&&\times\left.
\sqrt{-\det\left(\frac{}{}\eta_{ab}+\la^2 g_{YM}^2\prt_aX^i(Q^{-1})_{ij}\prt_bX^j
+\la (F_{ab}+\frac{1}{T_2}T^A_{ab})+\frac{1}{T_2}T^S_{ab}\right)} \right)\,\,.\nonumber \eeqa 
The matrices $Q^{ij}$, $T^S_{ab}$, $T^A_{ab}$ are  \beqa
Q^{ij}&=&I\delta^{ij}- \frac{g_{YM}^2}{T_2}[X^i,T] [X^j,T]\,,\qquad \det(Q)\,=\,1+\frac{g_{YM}^2}{T_2}[X^i,T][T,X^i]
\,\,,\labell{mq}\\
T^S_{ab}&=& D_aTD_bT+\frac{g_{YM}^2}{T_2}D_aT[X^i,T](Q^{-1})_{ij}[X^j,T]
D_bT\,,\nonumber\\
T^A_{ab}&=&i  g_{YM}^2\prt_aX^i(Q^{-1})_{ij}[X^j,T]D_bT-i  g_{YM}^2D_aT[X^i,T](Q^{-1})_{ij}
\prt_bX^j\,\,. \nonumber
\eeqa 
Here  the transverse scalars in \cite{Garousi:2000tr} are normalized as  $\Phi^i=g_{YM}\la X^i$  where $g_{YM}$ is the 3-dimensional Yang-Mills coupling constant, \ie $\lambda^2T_2=1/g^2_{YM}$, and a factor of $\sqrt{\la T_2}$ has been absorbed into the tachyon field. The tachyon potential is then a function of $T^2/(\la T_2)$. The trace in the action 
is completely symmetric between all  matrices
 $F_{ab},\prt X^i,D_aT, [X^i,T]$ and  individual
$T$ of the tachyon potential.
Hence, $(Q^{-1})_{ij}$ appears in symmetric form. Moreover, the symmetric trace makes the matrix $\eta_{ab}+\frac{1}{T_2}\prt_aX^i(Q^{-1})_{ij}\prt_bX^j+ \frac{1}{T_2}T^S_{ab}$ in the action to be symmetric and matrix $T^A_{ab}$ to be antisymmetric.  

Now we use the following de Wit-Herger-Samtleben duality transformation \cite{Iengo:2008cq}:
\beqa
-T_2\sqrt{-\phi\det(g_{ab}+\la F_{ab})}&\rightarrow &-T_2\sqrt{-\phi\det(g_{ab}+\frac{g_{YM}^2}{T_2}\frac{B'_aB'_b}{\phi})}+\frac{1}{2}\epsilon^{abc}B'_aF_{bc}\labell{iden}
\eeqa
for any scalar $\phi$, any symmetric matrix $g_{ab}$ and any antisymmetric matrix $F_{ab}$.
Using this duality in which $\phi=V\sqrt{\det(Q)}$ and $B'=VB$, the  action \reef{nonab} can be written in the following form:
 \beqa
S^{D\bar{D}}&=&-T_2\int
d^{3}\sigma \STr\left(V\sqrt{\det(Q)}\right.\labell{nonab1}\\
&&\times\left.
\sqrt{-\det\left(\eta_{ab}+\frac{1}{T_2}\prt_aX^i(Q^{-1})_{ij}\prt_bX^j+\frac{g_{YM}^2}{T_2}\frac{B_aB_b}{\det(Q)}
+\frac{1}{T_2}T^S_{ab}\right)} \right)\nonumber\\
&&+\frac{1}{2}\int
d^{3}\sigma \STr\left(\frac{}{}V\epsilon^{abc}B_a(F_{bc}+\frac{1}{T_2}T^A_{bc})\right)\,\,,\nonumber \eeqa 
where 
\beqa
 B_{a}=\pmatrix{B_{a}^{(1)}&0\cr
0&B_{a}^{(2)}}\nonumber
\eeqa
Near the unstable point of the tachyon potential one cat set $V\sim 1$. In the next section we are going to write an action for M2-brane-anti-M2-brane system around its unstable point that reduces to the above action around its unstable point under the Higgs mechanism \cite{Mukhi:2008ux}. 


\section{M2-brane-anti-M2-brane effective  action}

Using the prescription given in \cite{Mukhi:2008ux}, one  may expect that  effective action of the $M_2\bar{M}_2$  system to be reduced to the  effective action of $D_2\bar{D}_2$  system when $X^I_+$  takes a  large expectation value.  However,  the tachyon potential in the $M_2\bar{M}_2$  system may  not be related to  the  tachyon potential in the  $D_2\bar{D}_2$ system in this way since the $M_2\bar{M}_2$ action should describe the $D_2\bar{D}_2$ system at the  strong coupling limit. Moreover, it is expected that the tachyon potential at the strong coupling to be  totally different than    the tachyon potential at the weak coupling. However around their unstable point both potential are one. In this paper we are going to fix   the effective action of $M_2\bar{M}_2$ around its unstable point  by using the Higgs mechanism \cite{Mukhi:2008ux}.

The  prescription given in \cite{Mukhi:2008ux} has been used in \cite{Iengo:2008cq,Garousi:2008xn} to find a nonlinear action for multiple M2-branes. Following \cite{Iengo:2008cq}, the $M_2\bar{M}_2$  extension of  $S^{D\bar{D}}$ in \reef{nonab1} should have $SO(8)$ invariant terms $\tprt_aX^{I}(\tQ^{-1})_{IJ}\tprt_bX^{J}$ where $\tprt_aX^{I}$ and $\tQ_{IJ}$ should be defined to be invariant under the $B_a$ gauge transformation and   when $X^I_+=v\delta^{I10}$ where $v=g_{YM}$, they satisfy the boundary condition: 
\beqa
 \tprt_aX^{I}(\tQ^{-1})_{IJ}\tprt_bX^{J}\rightarrow \prt_aX^{i}(Q^{-1})_{ij}\prt_bX^{j}+v^2\frac{B^{}_aB^{}_b}{\det(Q)}
 \eeqa
This fixes $\tprt_aX^{I}$ to be \cite{Iengo:2008cq}
\beqa
\tprt_a X^{I}&=&\prt_aX^{I}-X^I_+B_a^{}-\left(\frac{X_+\inn X^{}}{X_+^2}\right)\prt_aX^I_+
\eeqa
where $X^2_+=X^I_+X^I_+$. This is invariant under the gauge transformation \reef{Bt}.
The boundary value of $\tQ_{IJ}$ is \cite{Iengo:2008cq}
\beqa
\tQ^{ij}=Q^{ij}\qquad,\qquad \tQ^{i10}=\tQ^{10i}=0\qquad,\qquad \tQ^{1010}=\det(Q)
\eeqa
At the boundary, one has $\det(\tQ)=(\det(Q))^2$.

 An ansatz  for $\tQ^{IJ}$ which is consistent with the above boundary condition may be 
 \beqa
 \tQ^{IJ}=a\delta^{IJ}+b\,M^{IK}M^{KJ}\nonumber
 \eeqa
 where $a,b$ are some $SO(8)$ invariants which can  be found from the above boundary condition, and
 \beqa
 M^{IJ}&\equiv&X^I_+[X^J,T]+X^J_+[T,X^I]
 \eeqa
 in which $X^I_+$ is singlet under $U(1)\times U(1)$ and
 \beqa X^I=\pmatrix{X^{I(1)}&0\cr
0&X^{I(2)}},\,\,T=\pmatrix{0&\tau\cr \tau^*&0} \labell{M12}\eeqa
Note that $\delta_B(M^{IJ})=0$ and consequently $\delta_B(\tQ^{IJ})=0$ if one assumes the tachyon to be  invariant under the $B_a$ gauge transformation. 
Imposing the boundary condition $\tQ^{ij}=Q^{ij}$ on the above ansatz, one finds
\beqa
\tQ^{IJ}&=&\delta^{IJ}+\frac{1}{T_2}M^{IK}M^{KJ} \eeqa
It also satisfies the boundary condition  $\tQ^{1010}=\det(Q)$. Using the relation between type IIA theory and M-theory, \ie $\ell_p=g_s^{1/3}\ell_s$, $T_2$ can be written in terms of 11-dimensional Plank length $\ell_p$ as $T_2=1/(2\pi)^2\ell_p^3$.

The matrices  $\tT^S_{ab}$ and $\tT^A_{ab}$ should be  determined  by forcing  them to be invariant  under global $SO(8)$ and under gauge transformation associated with $B_a$, and by imposing the boundary condition that at the boundary $X^I_+=v\delta^{I10}$ they should be reduced to those in \reef{mq}. The result is 
\beqa
\tT^S_{ab}&=&D_aT\left(\frac{1}{\sqrt{\det(\tQ)}}+\frac{1}{T_2}M^{IK}(\tQ^{-1})_{IJ}M^{KJ}\right)D_bT\nonumber\\
\tT^A_{ab}&=&i\tprt_aX^I(\tQ^{-1})_{IJ}M^{JK}D_bT X^K_+-iD_aTM^{IK}(\tQ^{-1})_{IJ}\tprt_b X^J X^K_+
\eeqa
Note  that the tachyon is invariant under the $B_a$ gauge transformation. 

Taking the above points, one finds that  the extension of the $D_2\bar{D}_2$ action \reef{nonab1} around its unstable point  to $M_2\bar{M}_2$  is then given by the  following action:
\beqa
&&\int d^3\sigma\,\STr \left(-T_2(\det(\tQ))^{1/4}
\sqrt{-\det\left(\eta_{ab}+\frac{1}{T_2}\tprt_aX^I(\tQ^{-1})_{IJ}\tprt_bX^J
+\frac{1}{T_2}\tT^S_{ab}\right)} \right.\nonumber\\
&&\left.+\frac{1}{2}\epsilon^{abc}\left(B_aF_{bc}-\frac{2i}{T_2}\tprt_aX^K\tprt_bX^I(\tQ^{-1})_{IJ}M^{JK}D_cT\right)\right)\nonumber 
\eeqa
where we have replaced $B_aX^K_+$ in the last line by the covariant expression $-\tprt_aX^K$. This action is manifestly invariant under global $SO(8)$, satisfies the Higgs mechanism   and is also invariant under gauge transformations associated  with gauge fields $A_a$ and $B_a$. The  symmetric trace   is  between the gauge invariants $\tprt_aX^I,D_aT,M^{IJ}$. 

The above action is not complete yet. Its tachyon potential is at its unstable point, \ie ${\cal V}\sim 1$, and it dose not reduce to the  action \reef{L1} at low energy and for  $T=0$. We assume that the tachyon potential appears in the action as an overall function as in tachyon DBI action.   The tachyon potential should be a function of only tachyon. The tachyon   is a  dimensionfull field, so   the tachyon potential  should be  a function of ${\cal V}\left(\frac{1}{T_2^{1/3}}T^2\right)$.
As we discussed before, this potential  is not expected to be reduced to the tachyon potential of the $D_2\bar{D}_2$ system under the Higgs mechanism.  To have consistency with  action \reef{L1}, one should add some extra terms to the above action \cite{Iengo:2008cq,Garousi:2008xn}. Hence, our proposal for the effective action of  $M_2\bar{M}_2$  is  the  following:
\beqa
&&\int d^3\sigma\,\STr \left({\cal V}\left[-T_2(\det(\tQ))^{1/4}
\sqrt{-\det\left(\eta_{ab}+\frac{1}{T_2}\tprt_aX^I(\tQ^{-1})_{IJ}\tprt_bX^J
+\frac{1}{T_2}\tT^S_{ab}\right)} \right.\right.\nonumber\\
&&\left.\left.+\frac{1}{2}\epsilon^{abc}\left(B_aF_{bc}-\frac{2i}{T_2}\tprt_aX^K\tprt_bX^I(\tQ^{-1})_{IJ}M^{JK}D_cT\right)\right.\right.\labell{L22}\\
&&\left.\left.+(\prt_aX^I_--\Tr(B_aX^I))\prt^aX^I_+-\Tr\left(\frac{X_+\inn X}{X^2_+}\hD_aX^I\prt^aX^I_+-\frac{1}{2}\left(\frac{X_+\inn X}{X^2_+}\right)^2\prt_aX^I_+\prt^aX^I_+\right)\right]\right)\nonumber 
\eeqa
 The last line in above action has been   added to have consistency at low energy and for $T=0$ with the action \reef{L1} for gauge group $U(1)\times U(1)$ \cite{Iengo:2008cq,Garousi:2008xn}. This action is manifestly invariant under global $SO(8)$  and is also invariant under gauge transformations associated  with gauge fields $A_a$ and $B_a$. 
 The  symmetric trace  in the first two lines is  between the gauge invariants $\tprt_aX^I,D_aT,M^{IJ}$ and individual $T$ of the tachyon potential, and in the last line it is only over the tachyon potential.

Let us now compare the two actions \reef{L22} and \reef{nonab1} around their unstable points where ${\cal V}\sim 1\sim V$. Action \reef{L22} gives the equation of motion for $X^I_-$ to be $\prt_a\prt^a X^{I}_+=0$. If one of the scalars $X^{I}_+$ takes large expectation value, \ie $X^{I}_+=v\delta^{I10}$, then $\tprt_aX^{i}=\prt_aX^{i}$,  $\tprt_aX^{10}=\prt_aX^{10}-vB_a^{}$ and $X_+\inn\prt_aX^{}=v\prt_aX^{10}$. Fixing  the gauge symmetry \reef{Bt}  by setting $X^{10}=0$,  one then recovers the $D_2\bar{D}_2$ action \reef{nonab1}. On the other hand, if the shift symmetry  $X^I_-\rightarrow X^I_-+c^I$ is gauged as in \cite{Bandres:2008kj,Ezhuthachan:2008ch} by introducing a new field $C_a^I$ and writing $\prt_aX^I_-$ as $\prt_aX^I_--C_a^I$, then equation of motion for the new field  gives $\prt_a X^{I}_+=0$ which has only constant solution $X^{I}_+=v^I$. Using the $SO(8)$ symmetry, one can write it as $X^{I}_+=v\delta^{I10}$. Then the $M_2\bar{M}_2$ theory \reef{L22} would be classically equivalent to the $D_2\bar{D}_2$ theory \reef{nonab1}. 
 

The $M_2\bar{M}_2$ action  \reef{L22}, for constant $X_+^I$, are written almost entirely in terms of covariant derivative of the scalars/tachyon and 3-bracket $M^{IJ}$. As pointed out in \cite{Alishahiha:2008rs}, one expects this part of the action which has no dependency on $X_+^I$  to be relevant to the theories beyond the Lorentzian-signature that we have considered here.  

{\bf Acknowledgments}:   This work is supported by Ferdowsi University of Mashhad under grant p/757(88/10/26). 


\begin{thebibliography}{99}

\bibitem{Schwarz:2004yj}
  J.~H.~Schwarz,
  JHEP {\bf 0411}, 078 (2004)
  arXiv:hep-th/0411077.

  A.~Basu and J.~A.~Harvey,
  Nucl.\ Phys.\  B {\bf 713}, 136 (2005)
  arXiv:hep-th/0412310.

\bibitem{Bagger:2007vi}
  J.~Bagger and N.~Lambert,
  JHEP {\bf 0802}, 105 (2008)
  arXiv:0712.3738.

  J.~Bagger and N.~Lambert,
  Phys.\ Rev.\  D {\bf 77}, 065008 (2008)
  arXiv:0711.0955.

  J.~Bagger and N.~Lambert,
  Phys.\ Rev.\  D {\bf 75}, 045020 (2007)
  arXiv:hep-th/0611108.
\bibitem{Gustavsson:2007vu}
  A.~Gustavsson,
  arXiv:0709.1260 .

  A.~Gustavsson,
  JHEP {\bf 0804}, 083 (2008)
  arXiv:0802.3456.
\bibitem{Lambert:2008et}
  N.~Lambert and D.~Tong,
  arXiv:0804.1114;
  J.~Distler, S.~Mukhi, C.~Papageorgakis and M.~Van Raamsdonk,
  JHEP {\bf 0805}, 038 (2008)
  arXiv:0804.1256;
 C.~Krishnan and C.~Maccaferri, 
 JHEP {bf 0807}, 005 (2008) 
 [arXiv:0805.3125 [hep-th]]. 


\bibitem{Aharony:2008ug}
  O.~Aharony, O.~Bergman, D.~L.~Jafferis and J.~Maldacena,
  arXiv:0806.1218 .
\bibitem{Gomis:2008uv}
  J.~Gomis, G.~Milanesi and J.~G.~Russo,
  JHEP {\bf 0806}, 075 (2008)
  arXiv:0805.1012.
  S.~Benvenuti, D.~Rodriguez-Gomez, E.~Tonni and H.~Verlinde,
  arXiv:0805.1087 .
  P.~M.~Ho, Y.~Imamura and Y.~Matsuo,
  JHEP {\bf 0807}, 003 (2008)
  arXiv:0805.1202.
\bibitem{Cecotti:2008qs}
  S.~Cecotti and A.~Sen,
  arXiv:0806.1990 [hep-th];
  P.~de Medeiros, J.~M.~Figueroa-O'Farrill and E.~Mendez-Escobar,
  arXiv:0806.3242 [hep-th];
  M.~Ali-Akbari, M.~M.~Sheikh-Jabbari and J.~Simon,
  arXiv:0807.1570 [hep-th];
  H.~Verlinde,
  arXiv:0807.2121 [hep-th].




  
\bibitem{Iengo:2008cq}
  R.~Iengo and J.~G.~Russo,
  arXiv:0808.2473 [hep-th].
\bibitem{Garousi:2008xn}
  M.~R.~Garousi,
  arXiv:0809.0985 [hep-th].

 
\bibitem{Li:2008ya}
  T.~Li, Y.~Liu and D.~Xie,
  arXiv:0807.1183 [hep-th].
\bibitem{Alishahiha:2008rs}
  M.~Alishahiha and S.~Mukhi,
  arXiv:0808.3067 [hep-th].


 
\bibitem{Kluson:2008nw} 
 J.~Kluson, 
 arXiv:0807.4054 [hep-th]. 
 
  
\bibitem{Mukhi:2008ux}
  S.~Mukhi and C.~Papageorgakis,
  JHEP {\bf 0805}, 085 (2008)
  [arXiv:0803.3218 [hep-th]].

  
  
\bibitem{Garousi:2004rd}
  M.~R.~Garousi,
  JHEP {\bf 0501}, 029 (2005)
  [arXiv:hep-th/0411222].
\bibitem{Garousi:2007fn}
  M.~R.~Garousi,
  JHEP {\bf 0712}, 089 (2007)
  [arXiv:0710.5469 [hep-th]].





\bibitem{Sen:1999md}
  A.~Sen,
  JHEP {\bf 9910}, 008 (1999)
  [arXiv:hep-th/9909062].
\bibitem{Garousi:2000tr}
  M.~R.~Garousi,
  Nucl.\ Phys.\  B {\bf 584}, 284 (2000)
  [arXiv:hep-th/0003122].
\bibitem{Bergshoeff:2000dq}
  E.~A.~Bergshoeff, M.~de Roo, T.~C.~de Wit, E.~Eyras and S.~Panda,
  JHEP {\bf 0005}, 009 (2000)
  [arXiv:hep-th/0003221].
\bibitem{Kluson:2000iy}
  J.~Kluson,
  Phys.\ Rev.\  D {\bf 62}, 126003 (2000)
  [arXiv:hep-th/0004106].
 

\bibitem{Garousi:2002wq}
  M.~R.~Garousi,
  Nucl.\ Phys.\  B {\bf 647}, 117 (2002)
  [arXiv:hep-th/0209068];
  M.~R.~Garousi,
  JHEP {\bf 0304}, 027 (2003)
  [arXiv:hep-th/0303239];
  M.~R.~Garousi,
  JHEP {\bf 0305}, 058 (2003)
  [arXiv:hep-th/0304145];
  M.~R.~Garousi,
  JHEP {\bf 0312}, 036 (2003)
  [arXiv:hep-th/0307197].
\bibitem{BitaghsirFadafan:2006cj}
  K.~Bitaghsir-Fadafan and M.~R.~Garousi,
  Nucl.\ Phys.\  B {\bf 760}, 197 (2007)
  [arXiv:hep-th/0607249];
  M.~R.~Garousi and E.~Hatefi,
  Nucl.\ Phys.\  B {\bf 800}, 502 (2008)
  [arXiv:0710.5875 [hep-th]].
\bibitem{Sen:2003tm}
  A.~Sen,
  Phys.\ Rev.\  D {\bf 68}, 066008 (2003)
  [arXiv:hep-th/0303057].
\bibitem{Bandres:2008kj}
  M.~A.~Bandres, A.~E.~Lipstein and J.~H.~Schwarz,
  arXiv:0806.0054 .
  
  J.~Gomis, D.~Rodriguez-Gomez, M.~Van Raamsdonk and H.~Verlinde,
  arXiv:0806.0738.
  
\bibitem{Ezhuthachan:2008ch}
  B.~Ezhuthachan, S.~Mukhi and C.~Papageorgakis,
  JHEP {\bf 0807}, 041 (2008)
  [arXiv:0806.1639 [hep-th]].





\end{thebibliography}
\end{document}